\documentclass[10pt]{iopart}
\usepackage{graphicx}
\usepackage{epsf}
\usepackage{latexsym}

\begin{document}

\title[Structural Transitions in A Crystalline Bilayer]
{Structural Transitions in A Crystalline Bilayer : The Case of Lennard Jones and Gaussian Core Models}
\author{Tamoghna Das, Surajit Sengupta}
 
\address{S. N. Bose National Center for Basic Sciences, 
Block-JD, Sector III, Salt Lake, Kolkata - 700098, India} 

\author{Subhasis Sinha}
\address{Indian Institute of Science, Education and Research Kolkata, 
Block-HC, Sector-III, Salt Lake, Kolkata - 700106, India}

\ead{tamoghna@bose.res.in}
\begin{abstract}
We study structural transitions in a system of interacting particles arranged 
as a crystalline bilayer, as a function of the density $\rho$ and the distance 
$d$ between the layers. As $d$ is decreased a sequence of transitions 
involving triangular, rhombic, square and centered rectangular lattices is 
observed. The sequence of phases and the order of transitions depends on the 
nature of interactions.
\end{abstract}

\section{Introduction}
Structural transitions in solids may be caused by various external parameters 
such as temperature, pressure, stress, electrical and magnetic fields \cite{stress.trans,field.trans}etc. Confinement and dimensional reduction can also lead to structural transitions especially in soft solids like colloidal suspensions. Colloidal solids are especially suited to modification and manipulation using a variety of means such as structural confinement \cite{str.conf}, laser-induced phase transitions \cite{lif}, shear \cite{shear}, static \cite{abhishek.prl} and 
dynamic \cite{ankush.prb} external fields.

Confined colloids kept in a thin wedge geometry of two optically-flat quartz 
glass plates, exhibit a sequence of structural transitions : 
$n\triangle \rightarrow (n+1)\Box \rightarrow (n+1)\triangle$ with 
increasing wedge height, where $n$ is the number of layers and $\triangle$ and 
$\Box$ corresponds to layers of triangular($p6$) and square($p4m$) symmetry 
respectively \cite{trans.exp1,trans.exp2}. The full equilibrium phase diagram of such a system has been studied analytically as well as using extensive computer simulations\cite{trans.simu1,trans.simu2}. The transitions are usually first order, though 
continuous transitions via a layer buckling mechanism \cite{buckling} has also
 been predicted and observed. 

In this paper, we explore another way in which structural transitions may be 
induced in a colloidal solid. Consider a crystalline bilayer separated by a 
distance $d$ between the layers. Each of these layers is held in place by 
individual trapping potentials, set up, for example using laser tweezers
\cite{laser.trap}. The strong trapping potential ensures that out of layer 
fluctuations are typically unimportant. We investigate the stability of the 
bilayer as the distance $d$ is decreased. We show that $d$ behaves as a 
controlling parameter and induces a rich sequence of transitions involving a 
variety of two-dimensional Bravais lattices. The exact sequence of transitions 
crucially depends on the nature of interactions. In this paper we study two 
kinds of model solids (a) the generic Lennard Jones (LJ) \cite{2dLJ.phasedia} 
solid and (b) the soft Gaussian core model (GCM) \cite{GCM.phasedia} 
appropriate  for suspensions of globular polymers. Our main results are as follows. For the LJ system, we obtain at temperature $T = 0$ two independent triangular (TRN) crystalline layers for large $d$. As $d$ is reduced, the system 
undergoes a {\em first order} transition to a staggered square (SQR) solid. 
As $d$ is further reduced, this square solid becomes, first, a centered 
rectangular (CR) and finally again a triangular solid as $d$ is decreases to 
zero and the layers merge. The final two transitions are continuous. 

In contrast for the GCM, {\em all} the transitions are continuous. 
As $d$ reduces, the TRN solid transforms to a SQR solid continuously through a 
sequences of rhombic (RMB) lattices. The SQR solid subsequently transforms back
 to the TRN solid for small values of $d$ again continuously but this time it 
uses a sequence of CR lattices with intermediate aspect ratios. The progression
 of phases seen correspond roughly with those seen in the extensive literature 
on classical interacting bilayer Wigner crystals \cite{2lWignercrystal}, though the sequence of phases and the nature of transitions are different.  

The rest of the paper is organized as below. In the next section, we describe 
the bilayer system in detail, introducing the order parameters for the 
transition and state the interatomic potentials used. In Section 3,we give 
the results for the zero temperature energy minimization. This is followed in 
Section 4 by a full normal mode analysis investigating the stability of the 
ground states obtained in Section 3 and the nature of the transition. In 
Section 5, we present results of finite temperature Monte Carlo simulations. 
We discuss our results and their implications and conclude in Section 6. 

\section{Model System}
\begin{figure}[h]
\begin{center}
\includegraphics[width=3.0cm]{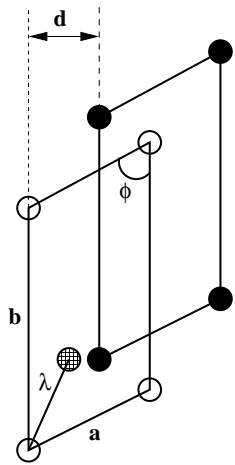}
\caption{
A schematic diagram of the model bilayer solid explaining the structural
parameters. The crystal structure on each of the layers, denoted by filled 
and open circles, are identical but staggered by the amount $\lambda$ along
the diagonal. The lattice parameetrs are $a$ and $b$ and the apex angle is 
$\phi$. The half filled circle is the projection of a lattice point (filled 
circle) on the top layer to the bottom one. 
}
 \label{fig1}
\end{center}
\end{figure}

Consider a system of $2N$ particles arranged as two parallel two-dimesional 
crystalline layers of $N$ particles each (see Fig.(\ref{fig1})). The crystal 
structure of each of these layers may be assumed to be a general 
two-dimensional oblique ($p2$) lattice. Each particle interacts with all other 
particles via isotropic and pair-wise interacting potentials. The position 
vector for the $i$-th particle of this lattice can thus  be expressed as
\begin{eqnarray}
\vec{r}_{i} &=& (m + \lambda)\vec{a} + (n + \lambda)\vec{b} + d\hat{z} \nonumber \\
                    &=&((m + \lambda)a + (n + \lambda)b{}\cos\phi){}\hat{x} +{}(n + \lambda){}b{}\sin\phi{}\hat{y} + d{ }\hat{z}
\label{eqn1}                    
\end{eqnarray}
where $m,n = 0,1,2,\cdots$, $\vec{a}$ and $\vec{b}$ are two in-plane basis 
vectors, $\phi$ is the angle between these two basis vectors,$d$ is the 
interlayer separation and $\lambda$ is a shift between the center of masses of 
two layers (Fig.(\ref{fig1})). Therefore to specify our model completely, 
knowledge of these five variables, $a, b, \phi, d, \lambda$ is sufficient. The 
last variable $\lambda$ ensures that for small values of $d$, the particles 
from different layers do not overlap. The particles are not allowed to fluctuate out 
of the layers and particle exchange between the layers is prohibited. Each 
layer is therefore considered to be strongly confined in the Z-direction while 
they are allowed complete freedom in the X,Y- plane.

Let us first imagine the possible physical scenario as the layers are brought 
close to each other starting from a large inter-layer separation. When the 
layers are well apart they exist as two independent mono-layers and they show 
TRN symmetry, the minimum energy configuration for the two-dimensional 
crystalline system for the interaction potentials considered by us. As the 
inter-layer separation, $d$, between these two layers starts decreasing, the 
system passes through a series of structural transitions which may involve RMB 
and SQR phases. For even smaller values of $d$, two layers start merging into 
one and the TRN symmetry is regained. Transformation between a SRQ and 
a TRN phase may be accomplished, in general, by either (i) shear i.e. change in 
the angle between two in-plane basis vectors producing an intermediate RMB 
structure or (ii) change in the aspect ratio ($b/a$) which produces a CR 
lattice. Both the RMB and the CR lattices being less symmetric have TRN and 
SQR phases as limiting cases.

It is therefore clear that we need to introduce two order parameters \cite{group.rep} in order to describe completely the phase transitions of our model system.
 First, the bond angle order parameter $\psi = \cos\phi$ which is $0$ when 
system takes SQR symmetry ($\phi=90^\circ$) and non-zero otherwise.
 The second order parameter $\xi$ is related to the aspect ratio $b/a$ as 
$\xi = (b/a - 1)/ (\sqrt{3} - 1)$ which varies from $0$ in the SQR to a 
non-zero value in the CR phase. Note that the highly symmetric TRN phase is 
described both by ($\psi = 0.5, \xi = 0$) and ($\psi = 0.0, \xi = 1$).
Finally, if $\epsilon_{ij}, i,j = x,y$ is the two dimensional strain tensor, 
the shear, $e_3 = \epsilon_{xy}$, and deviatoric, $e_2 = \epsilon_{xx} - 
\epsilon_{yy}$, strains are related to $\theta$ and $\xi$ as,
\begin{eqnarray}
\tan{}\phi &=& \epsilon_3/(1 - \epsilon_1) \nonumber \\
\phantom{xxx}{}\xi &=& \frac{2(1 + \epsilon_2) + \epsilon_1}{2(1 - \epsilon_2) + \epsilon_1}
\label{eqn2}
\end{eqnarray}
 
We have studied phase transitions in the bilayer system for two different model potentials. The Lennard Jones potential :
\begin{eqnarray}
U &=& \sum_{i \neq j} V_{LJ}(r_{ij}) \nonumber \\
              &=& 4 \epsilon \biggl[\biggl(\frac{\sigma}{r_{ij}}\biggr)^{12} -\biggl(\frac{\sigma}{r_{ij}}\biggr)^6 \biggr]
\label{eqn3}
\end{eqnarray}
has been used extensively in the past as a generic model which includes both 
long range attractive and short range repulsive interactions.  In Eq.
(\ref{eqn3}), $r_{ij} = | \vec{r}_i - \vec{r}_j |$, the distance between the 
$i$ and $j$-th particle. An intrinsic length scale $r_{min} = 2^{1/6} \sigma$
corresponding to the minimum of $V_{LJ}$ may be associated with this 
potential. The nearest neighbour distance between particles is close to this 
value throughout. We use reduced units for LJ potential throughout the paper 
defining lengths in units of $\sigma$ and energy in units of $\epsilon$. It 
follows that the densities are in units of $\sigma^{- 3}$ and temperatures in 
units of $\epsilon/k_B$, where $k_B$ is the Boltzmann factor.

On the other hand, {\it Gaussian core} potential has been used specifically to model soft solids.
\begin{eqnarray}
U &=& \sum_{i \neq j} V_{GCM}(r_{ij}) \nonumber \\
&=& \sum_{i \neq j} V_0 exp[- (r_{ij}/l)^2],\phantom{xxx}V_0,l > 0
\label{eqn4}
\end{eqnarray}
$V_0$ and $l$ acts as the energy and length-scales for this potential and 
$r_{ij}$ is same as it was for the previous equation. The GCM is interesting 
because, firstly, the potential is soft and is known to display behaviour 
similar to that of real polymeric solids. Also many of its properties are 
well known and tested especially because at very low densities it reduces
to the hard disk model which is widely used for modelling colloidal solids.
Due to the purely repulsive nature of this potential, it does not 
have any preferred nearest neighbour distance which is determined in this case 
by the density. This potential also possesses an interesting duality property
\cite{GCM.duality} such that high ($\rho_>$) and low ($\rho_<$) density 
properties are related to each other by
\begin{equation}
\rho_> \rho_< = \pi^{-D}
\label{eqn5}
\end{equation}
where $D$ is the dimensionality. It is therefore sufficient to confine our 
studies in the range $0 < \rho < \rho_{f.p.}$ where the fixed point density 
$\rho_{f.p.} = \pi^{-1} \simeq 0.32$ in two-dimensions.

\section{Zero Temperature Calculation}

In this section, we determine the ground states of our system in the space of 
the order parameters, $\phi$ and $\xi$, keeping the layer separation $d$ 
and the density $\rho$ as external parameters. Below we discuss our results 
for the LJ solid and the GCM one after the other.

\subsection{Lennard-Jones potential}
We begin with $2N$ particles divided into two layers each arranged in a 
TRN lattice of $N$ particles. For fixed layer separation $d$, and 
the lattice parameter $a$ set by the density $\rho$,
we minimize the total energy $U$ with respect to $\lambda$ and $b$. The 
resulting order parameter $\xi$ remains constant at $1$ -- the value 
appropriate for the TRN lattice. For the lattice sums needed to calculate 
the minimized energy we have set a cut-off radius, $r_c = 2.5 \sigma$.

\begin{figure}[h]
\begin{center}
\includegraphics[width=12.0cm]{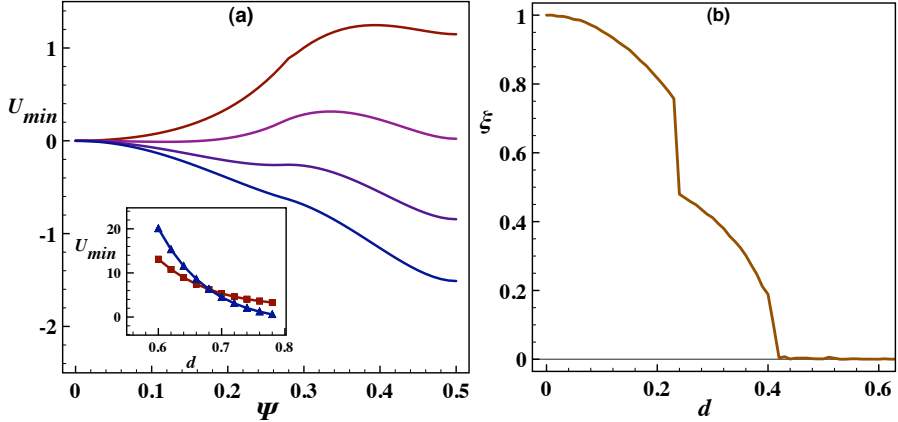}
\caption{(a) Plots of the total energy as a function of $\psi$ at 
$\rho = 1.2$ and (top to bottom) $d = 0.66,0.68,0.70$and $0.72$ showing the
SQR to TRN transition for the LJ system. (inset) The energies of the SQR 
(filled square) and TRN (filled triangles) plotted as a function of $d$ 
are seen to cross with a change in slope as expected of a first order 
transition. The fact that this transition is first order has also been
verified by a normal mode analysis. (b) Plot of the order parameter $\xi$ 
as a function of $d$ at the same density showing the two continuous transitions
SQR $\to$ CR $\to$ TRN as $d \to 0$.}
\label{fig2}
\end{center}
\end{figure}

In Fig.(\ref{fig2}a) we have plotted minimized energy $U_{min}$ as a function 
of the order parameter $\psi$ for various values of $d$. For large inter-layer 
separation, bond angle order parameter at the minimized energy
($U_{min}$) shows a non-zero value, $\psi = \cos \phi = 0.5$, which indicates 
TRN symmetry ($\phi = 60^\circ$). The minimum in $\lambda$ is very shallow 
indicating two triangular lattices prefer to remain independent. As $d$ is 
decreased two layers fall into registry and for further decrease of $d$ a 
second minimum at $\psi = 0$ develops corresponding to the SQR phase 
($\phi = 90^\circ$). A plot of the energy of the SQR and TRN phase
(Fig.(\ref{fig2}a) inset) shows a first order transition at $d = 0.684$.  
This fact is confirmed by a normal-mode analysis presented in the next section.

At even smaller values of $d$, the square solid begins to deform continuously 
by changing the aspect ratio (or order parameter $\xi$) at fixed $\phi$. In 
Fig.(\ref{fig2}b) we have plotted the minimized value of $\xi$ vs. $d$. From 
the plot it is obvious that there are two distinct continuous transitions 
taking the solid from SQR at $\xi = 0$ to an eventual TRN phase at $\xi = 1$ 
via an intermediate CR lattice of $\xi = 0.5$.

\subsection{Gaussian Core Model}
\begin{figure}[h]
\begin{center}
\includegraphics[width=12.0cm]{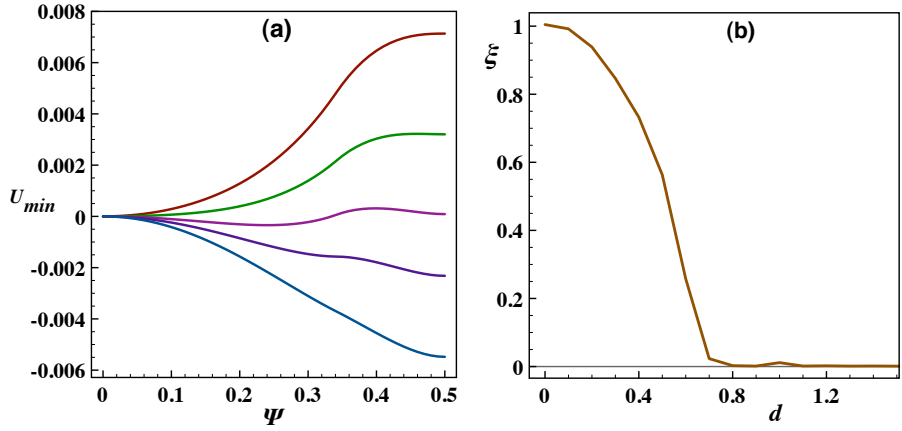}
\end{center}
\caption{(a) Plot of the total energy in the GCM for $\rho = 0.2$ and (top to 
bottom) $d = 1.4, 1.5, 1.6, 1.7$ and $1.9$ as a function of $\psi$. Note the 
appearance of the minimum at $\psi = 0.5$ corresponding to the TRN lattice as 
$d$ increases. (b) The continuous transition from the SQR to the TRN lattice 
as $d \to 0$.}
\label{fig3}
\end{figure}
Due to the purely repulsive nature of the Gaussian core potential, it does not 
have any preferred nearest neighbour distance and at zero stress the solid 
would disintegrate. We have carried out all minimizations for a system with 
fixed reduced density, $\rho = 0.2$. We have plotted the minimized energy 
$U_{min}$ as a function of the bond angle order parameter $\phi$ for various 
values of $d$.  In Fig.(\ref{fig3}a) in contrast to the Lennard-Jones solid we 
now obtain continuous transitions from TRN to SQR through a set of RMB phases 
and back to TRN at $d = 0$ through a set of CR phases.
 
\subsection{T=0 Phase Diagram}
\begin{figure}[h]
\begin{center}
\includegraphics[width=12.0cm]{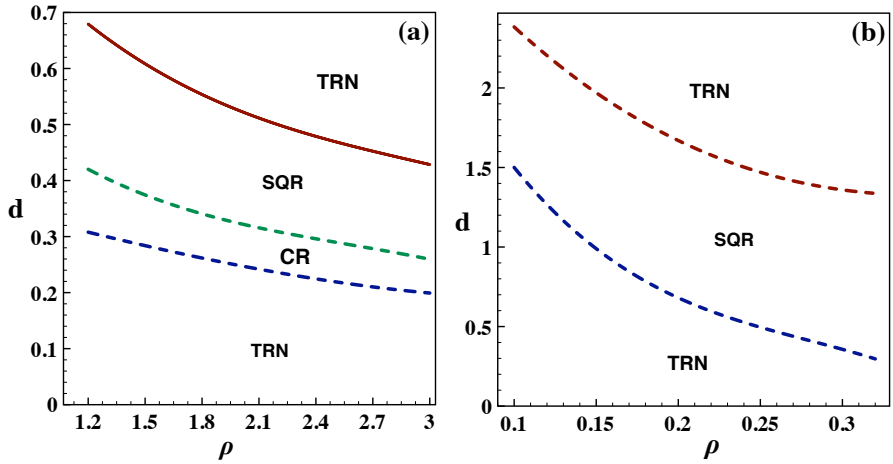}
\end{center}
\caption{Zero temperature phase diagrams in the $\rho - d$ plane for 
the LJ (a) and GCM (b) systems. The various phases are marked. First order 
transitions are shown by solid and continuous transitions by dashed lines. 
} 
\label{fig4}
\end{figure}
Our results concerning the various ground states and structural transitions in 
both the LJ and GCM systems in the $\rho$ - $d$ plane has been shown in 
the zero temperature phase diagrams, Fig.(\ref{fig4}a) and (b) respectively.
Note that for both the systems the triangular phase is stable at all 
$\rho$ for both very large $d$ and $d=0$ where the system becomes effectively 
two dimensional.  

\section{Normal Mode Analysis}

To further elucidate the nature of the structural transitions in the two 
systems, we have undertaken a normal mode analysis \cite{ashcroft.book} of the TRN and 
SQR solids obtained for each of the two interactions. 

Let $\vec{u}(\vec{r}_i)$ be the displacement of the $i$-th particle from its 
equilibrium position $\vec{r}_i$. Within harmmonic approximation, now the 
potential can be written as
\begin{equation}
U_{harm} = \frac{1}{2}\sum_{i,j} \vec{u}(\vec{r}_i) \mathcal{D}(\vec{r}_i - \vec{r}_j) \vec{u}(\vec{r}_j)
\label{eqn6}
\end{equation}
\begin{eqnarray}
\mathcal{D}(\vec{r}_i - \vec{r}_j) &=& D_{\mu \nu}(\vec{r}_i - \vec{r}_j) \nonumber \\
&=& \delta_{\vec{r}_i \vec{r}_j}\sum_k V_{\mu \nu}(\vec{r}_i - \vec{r}_k) - V_{\mu \nu}(\vec{r}_i - \vec{r}_j)
\label{eqn7}
\end{eqnarray}
where $V_{\mu \nu}(\vec{r}) = \partial^2 V/\partial r_\mu \partial r_\nu$. 
We have $2N$ equations of motion, one for each of the three components of the 
$N$ particles, since we have already restricted fluctuation in the 
$z$-direction.
\begin{equation}
\ddot{\vec{u}}(\vec{r}_i) = - \sum_j \mathcal{D}(\vec{r}_i - \vec{r}_j) \vec{u}(\vec{r}_j)
\label{eqn8}
\end{equation}
We seek solutions to the equations of motion in the form of simple plane waves :
\begin{equation}
\vec{u}(\vec{r},t) = \vec{\epsilon}\phantom{x}exp[i(\vec{k}. \vec{r} - \omega t)]
\label{eqn9}
\end{equation}
Here $\vec{\epsilon}$ is the {\it polarization vector} of the normal mode. The 
{\it Born-vonKarman} periodic boundary conditions restricts the wave vector 
$\vec{k}$ to a single primitive cell of the reciprocal lattice vector, which 
is normally identified with the first Brillouin zone.

\begin{figure}[h]
\begin{center}
\includegraphics[width=14.0cm]{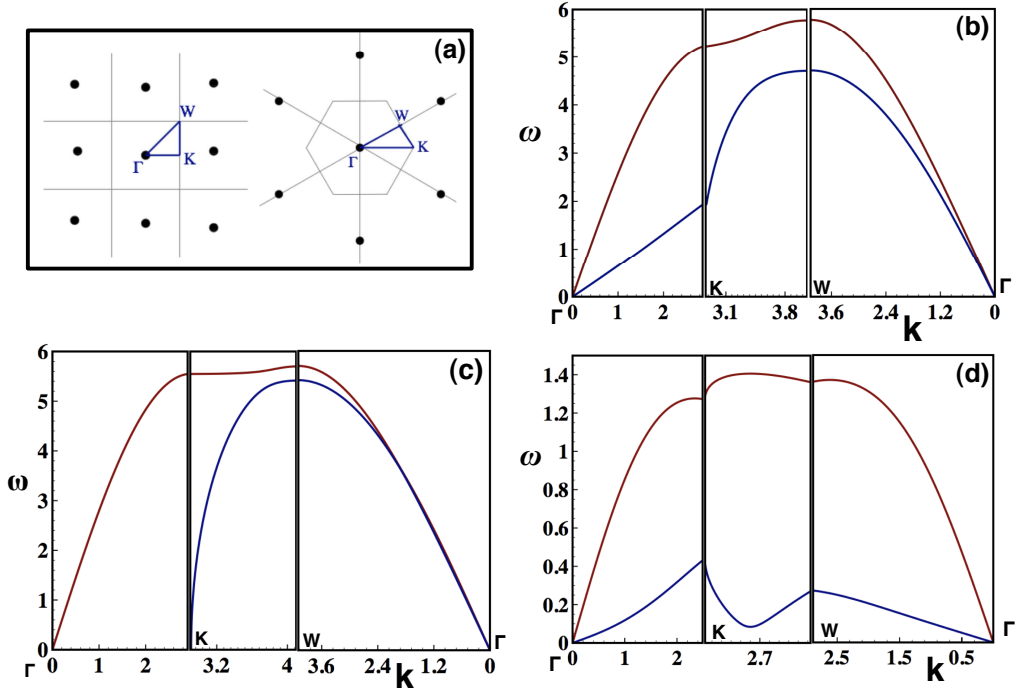}
\caption{Results for phonon dispersion curves : (a) diagram showing the 
high symmetry points in the SQR and TRN (reciprocal) lattices. (b) Phonon
dispersion for the stable SQR lattice at $\rho = 1.2$ and $d=d_c=0.684$ i.e.
the value of $d$ at SQR to TRN transition. Note that the dispersion curve
shows that the SQR phase is locally stable pointing to a first order transition.
(c) Phonon dispersion of the SQR solid at $\rho = 1.2$ and $d = 0.4$ showing 
an instability in the transverse acoustic branch. At this value of $d$ the CR 
solid is stable. (d) Dispersion curve for the TRN solid in the GCM for 
$\rho = 0.2$ and $d = 1.6$ showing the appearance of a non zero $k$ vector soft 
mode.}
\label{fig5}
\end{center}
\end{figure}
 
Substituting Eq.(\ref{eqn8}) into Eq.(\ref{eqn7}) we find a solution of the 
three-dimensional eigen value problem :
\begin{equation}
\omega^2 \vec{\epsilon} = \mathcal{D}(\vec{k}) \vec{\epsilon}
\label{eqn10}
\end{equation}
Here $\mathcal{D}(\vec{k})$, the {\it dynamical matrix}, is given by
\begin{equation}
\mathcal{D}(\vec{k}) = \sum_i \mathcal{D}(\vec{r}_i) e^{-i \vec{k}. \vec{r}_i}
\label{eqn11}
\end{equation}
Two solutions to Eq.(\ref{eqn9}) for each of the $N$ allowed values of 
$\vec{k}$ give us $2N$ normal modes. The reciprocal lattices for both SQR 
and TRN are known to have the same symmetry of the real space lattice. 
Exploiting this property, one finds the values of $\omega$ only for those 
$k$-values along the lines connecting the high symmetry points of the 
first Brillouin zone (Fig.(\ref{fig5}a)) thereby obtaining the dispersion 
curve $\omega$ vs $|\vec{k}|$ and mode structure of a given lattice.
For any stable equilibrium structure $\omega$ should always be non-negative 
definite.

\subsection{The Lennard Jones potential}

To show that the SQR $\rightleftharpoons$ TRN transition at large $d$ is 
indeed first order, we have obtained the dispersion curves for the metastable 
SQR phase for a value of $d$ slightly larger than the critical $d_c = 0.684$  
for the chosen $\rho = 1.2$. This is shown in Fig.(\ref{fig5}b). This indicates 
that the transition is first order with the possibility of co-existence. When 
$d$ is decreased further, $\psi$ at the minimized energy $U_{min}$ shows 
minima at $\psi = 0$ but the normal mode analysis shows us that the SQR 
structure can not be stable at this value of $d$ (Fig.(\ref{fig5}c)). Actually,
the two layers start merging into one, by changing the aspect ratio away from 
$b/a=1$. 

\subsection{The Gaussian Core Model}

For the Gaussian core model the scenario is quite different. For intermediate 
values of $d$, the SQR structure is seen to be unstable and the mode 
structure for the TRN structure exhibits {\em mode softening},
Fig.(\ref{fig5}d)), such that $\omega \to 0$ for $k \neq 0$. Examination of 
the deformation corresponding to this $\vec k$ shows that the SQR lattice 
becomes unstable to shear deformation at the zone boundary.  This mode 
softening therefore establishes the transition from TRN to SQR
transition to be continuous for the Gaussian core model.

The transition from SQR back to TRN at low $d$ is always continuous 
both for LJ and GCM as verified by our normal mode analysis.

\section{Finite Temperature results: }

We close our discussion on structural transitions in a bilayer crystal by 
briefly mentioning some of our results for the LJ case using Monte Carlo 
simulations \cite{frenkel.book}. A detailed calculation of the phase diagram
of the bilayer GCM in the temperature, $\rho$ and $d$ space using both Monte 
Carlo and classical mean field density functional theory, which is known to 
yeild particularly good results for this system, will be published elsewhere.  

\begin{figure}[h]
\begin{center}
\includegraphics[width=12.0cm]{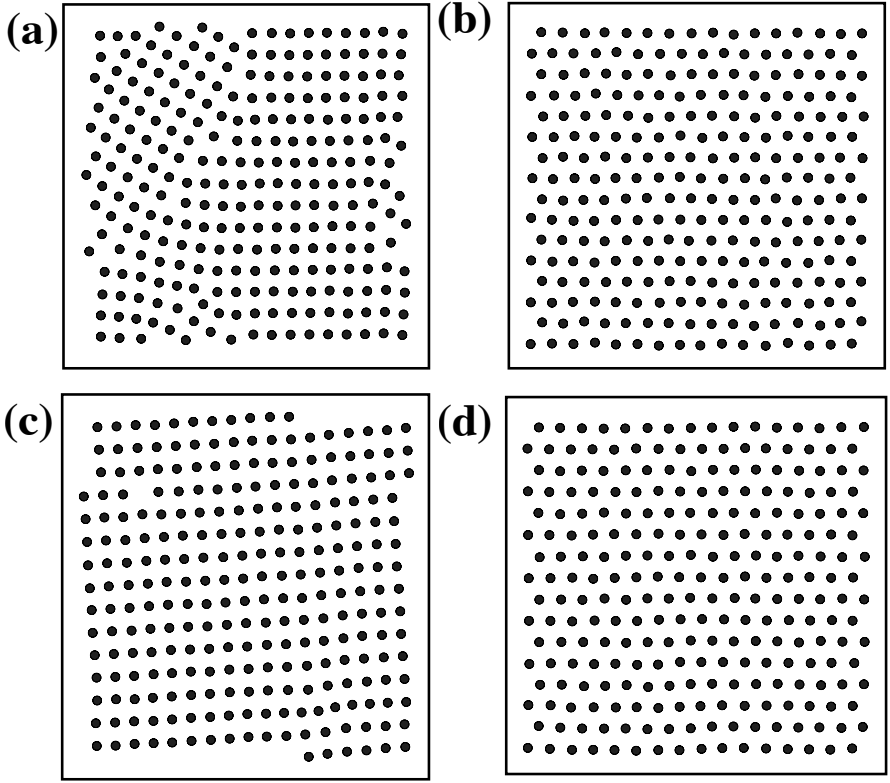}
\end{center}
\caption{Configurations from our Monte Carlo simulation in the LJ system  
for $T = 1.0$. (a) and (b) show the result of increasing $d$ from $d = 0.6$ to 
$d = 0.8$ at fixed $\rho = 1.20$. Note the first order SQR to TRN transition.
(c) and (d) show the corresponding result for pressure induced SQR to TRN 
transition as $\rho$ is increased from $1.3$ to $1.4$ at fixed $d = 0.6$. Our 
results are consistent with those seen in Fig.(\ref{fig4}a). } 
\label{fig6}
\end{figure}

The simulation is done using the usual Metropolis algorithm keeping total 
number$(N)$ of particles, volume$(V)$ and temperature$(T)$ fixed.
Periodic boundary conditions have been used for all directions except in the 
direction of stacking of the layers. For our purpose, we use a system of $512$ 
number of particles, temperature is fixed at $1.0$ and the volume is 
determined by the density, $\rho = 1.2$, used for simulation.
Starting from a large layer seperation $(d\gg0)$, we have observed the system 
change its symmetry from TRN to SQR with decreasing $d$. For large values of 
$d$ ($d > 0.8$), when the two layers are well seperated, each layer behaves independently
 of each other exhibiting the expected TRN structure. As $d$ is 
decreased, the two layers start interacting with each other. As a result, part 
of the system starts to transform into a SQR. Through this phase co-existence, 
the whole system transforms to a state where each layer shows SQR structure at 
$d = 0.6$. 

We have also explored the configurations of our model system in the 
temperature-density plane keeping the inter-layer separation fixed. For 
$d=0.6$, where we have already seen the SQR structure to be the ground state,
we increase $\rho$ to $1.3$. We again encounter a first order boundary and 
the system equilibrates to the SQR structure. 

The values of the critical $d$ and $\rho$ compare favourably with the 
$T=0$ phase diagram shown in Fig. (\ref{fig4}a). Similar scans at other 
temperature values over a large range of $\rho$ and $d$ have confirmed that 
it is not possible to induce a structural transition in this model by changing 
$T$. The phase boundaries shown in Fig.(\ref{fig4}) therefore extend vertically 
upto a melting temperature $T_m(\rho,d)$. This behavior is identical to that 
seen in the classical Wigner crystal and is probably an universal feature of 
any such crystalline bilayer. 

\section{Summary and Conclusion}

In this paper we have studied structural transitions in a bilayer crystal. We
have shown that the system has a rich phase diagram and shows a number of 
phase transitons. The identity of the phases and the nature of the transitions
depend on the interaction potential. On the other hand, some features of these 
transitions e.g. the overall topology of the phase diagram seems to be similar
and independent of the details of the interaction.

We believe that it may be easy to realize this soft matter system 
experimentally and study many of its interesting equilibrium and dynamic 
characteristics. For example, critical properties of the continuous 
structural transitions, especially, near the melting line and a detailed study of finite size scaling and crossover in these systems may be 
illuminating \cite{Chaikin}. We are also particularly interested in the dynamics of the structural transitions for both the first order and continuous cases. 
Quenches from the SQR to the TRN lattice in this system may be accomplished 
simply by changing $d$. How does the new phase form inside the parent? 
Is there a possibility of a martensitic transition\cite{NAZ1,NAZ2}? If so, then of what type?  We hope our work stimulates experiments designed to answer 
these questions in the near future.  

\section{Acknowledgement}

We acknowledge useful discussions with K. G. Ayappa, M. Rao and A. Paul.

\end{document}